# Elastomeric carbon nanotube circuits for local strain sensing

Hareem Maune, Marc Bockrath

*California Institute of Technology Mail Stop 128-95 Pasadena, CA 91125*

We use elastomeric polydimethylsiloxane substrates to strain single-walled carbon nanotubes and modulate their electronic properties, with the aim of developing flexible materials that can sense local strain. We demonstrate micron-scale nanotube devices that can be cycled repeatedly through strains as high as 20% while providing reproducible local strain transduction by via the device resistance. We also compress individual nanotubes, and find they undergo an undulatory distortion with a characteristic spatial period of 100-200 nm. The observed period can be understood by the mechanical properties of nanotubes and the substrate in conjunction with continuum elasticity theory. These could potentially be used to create superlattices within individual nanotubes, enabling novel devices and applications.

Electronics incorporated into plastics promise a wide array of low-cost or novel applications because of their light weight and mechanical flexibility or elasticity.[1-7] Up to now, most efforts have been concentrated on placing standard circuits on such substrates. Left largely unexplored is utilizing the substrates' capability to change shape to influence device behavior. Such devices may allow carbon nanotube or semiconductor nanowire circuits in which their electronic and optical properties intrinsically depend on the shape of the substrate. This could potentially enable a wide variety of applications such as nano-scale strain sensing, flow sensing in nanofluidics, and mechanically tunable optical devices, as well as producing "artificial skin" membranes with local pressure or shape sensitivity. Such skins may be important, for example, in robotics applications in which spatially resolved touch sensitivity is required, or for smart materials that can sense shape changes with high-resolution.[8,9]

As an initial demonstration of this principle, we have fabricated single-walled carbon nanotube devices, consisting of a thin film of nanotubes bridging gold electrodes, on elastomeric polydimethylsiloxane (PDMS) substrates. While monitoring the device resistance, we elongate the nanotubes an order of magnitude more than in previous experiments in which the transport properties were simultaneously observed.[10,11] We find that our devices can withstand up to ~20% strain while providing a measurable and repeatable change in resistance. In addition to potential applications, this opens the possibility of obtaining important insight into the fundamental properties of the nanotubes and engineering novel nanotube strain-based devices. We also demonstrate the compression of nanotubes, which produces a wavelike distortion due to a mechanical buckling instability that could potentially create strain-induced periodic quantum well structures.

Our devices are fabricated by evaporating Cr/Au electrodes on the PDMS substrate through a Riston[DuPont] foil shadow mask with features defined using optical lithography, as shown in Fig. 1(a).[12] A thin wire (~3 μm diameter) is placed across the mask in order to define a gap between the electrodes. We maintain the PDMS in a stretched state during the evaporation so that upon release, the metal is under compression and can be subsequently stretched without the metal breaking. Following the metal deposition, we deposit nanotubes from a suspension in dimethylformamide on to the electrodes, yielding a device as shown in the atomic force microscope (AFM) image in Fig. 1(b). The electrodes are typically bridged by isolated regions of nanotubes with sufficient density to provide a percolating pathway for charge carriers between the electrodes. Once we place the nanotubes on the electrodes, we measure the electrical transport characteristics through the attached source and drain electrodes.

To study the effects of strain on our devices, the entire PDMS substrate in placed under tensile strain using a clamp with a manual setscrew. The electrodes are capable of sustaining ~20% strain without breaking and maintain a low-impedance current path to the nanotubes, as previously reported by Lacour et al.[12] The substrate strain elongates the nanotubes, as shown in Fig. 1(c). The left panel of Fig. 1(c) shows an image of a nanotube on an unstretched substrate, while the right panel shows an image with the substrate stretched by ~20%. A direct measurement of the nanotube lengths before and after stretching shows a nanotube strain of ~15%. Examination of the electrode gap under an optical microscope enables us to determine the local strain $\sigma$ applied to the nanotubes. This local strain is typically 3-5 times larger than the overall substrate strain, due to the surface stress applied by the relatively rigid metal electrodes.

With the attached electrodes, we measure the changes in the tubes' transport properties due to changes in $\sigma$, which are an order of magnitude larger than in previous experiments studying individual nanotube transport properties under strain.[10,11] Fig. 2(a) shows current-voltage (*I-V*) curves for a representative device D1, consisting of ~10 nanotubes in parallel obtained with $\sigma$ = 0% and 10%. The device resistance *R* increases considerably from ~50 MΩ to ~1 GΩ under the applied strain. To investigate the reproducibility of the resistance changes with $\sigma$, we cycled $\sigma$ numerous times on a different device D2. Figure 2(b) shows resistance data vs. time taken from D2. During the data trace, $\sigma$ was first square-wave modulated between 0% and 10%, modulating the resistance between ~1.3 MΩ and ~1.4 MΩ. We then square-wave modulated $\sigma$ between 10% and 20%, modulating the resistance between ~1.5 MΩ and ~1.6 MΩ. Note that there is some hysteresis in the resistance values. This is likely due at least in part to hysteresis in the mechanical straining system and setscrew position. Allowing for the hysteresis, upon repeated straining and relaxation, the resistance values are reproducible and yield ~1% resistance modulation per % of substrate strain. The fractional resistance change Δ*R*/*R* is plotted vs. $\sigma$ in the inset to Fig 2(b). This demonstrates that multiple nanotube or nanotube network devices are piezoresistive and can withstand up to 20% strain without irreversible changes to their electronic properties.

To understand the $\sigma$ dependence of our devices' resistance, we assume that the applied strain changes the band gaps of the nanotubes,[10,11,13] producing changes in the resistance. We model the composition of the thin film networks in our experiment as nanotubes with random chirality, with most of the nanotubes in the measured diameters for HiPco-produced nanotubes in the range of 0.8-1.1 nm.[14,15] The band gap for each nanotube in the ensemble is computed using the relation

$$\Delta E_{gap} = \text{sgn}(2p+1)3t_0[(1+\nu)\,\sigma \cos^2(\phi) \cos(3\theta)], \qquad (3)$$

where $p$ depends on both the chiral indices and strain state of the nanotube and is $\pm 1$ for unstrained semiconducting nanotubes and 0 for unstrained metallic or small-bandgap tubes, $t_0$=2.66 eV, $\theta$ is the nanotube chiral angle, and $\nu$=0.2 is the Poisson ratio.[13] For each tube in the ensemble, we also include a random angle $\phi$ in relation to the elongation axis so that to lowest order in $\sigma$, its net strain is reduced from the local substrate strain by a factor $\cos^2(\phi)$.

Histograms of the bandgaps of the nanotube ensemble of 660 tubes for $\sigma$=0%, 10%, and 20% are shown in Fig. 3. Between $\sigma$=0% and 10%, we expect the resistance to increase. This should occur because the number of metallic or small-bandgap nanotubes is dramatically reduced, and these nanotubes initially contribute the least resistive current pathways to the thin-film.[16] Indeed we find such an increase for both D1 and D2. It is difficult to make quantitative predictions for the magnitude of the resistance changes in our experiment because we do not employ a gate electrode in our devices. As a result, the carrier concentration is not known precisely. Between $\sigma$=10% and 20% in our experiment we see a further increase in resistance for D2, although from the histogram corresponding to 20% strain, we might expect similar resistance to that at 10%. As it is unknown what the effects of such large strains are on transport properties, for example the carrier mobility, an accounting for the large strain regime will likely require further theoretical modeling. Measurements on individual nanotubes at these large strains are also highly desirable. Nevertheless, we have demonstrated that empirically, nanotube networks are capable of reproducible local strain transduction even under very large strains up to 20%. Arrays of these devices may enable "shape-aware" materials that sense local shape distortions in a spatially resolved manner.

Another unique aspect of our experimental approach is that we have the capability to readily place nanotubes under compressive strain. Figure 4(a) shows a topographic AFM image of a chemical vapor deposition grown nanotube[17] placed on an elongated PDMS substrate by contact transferring them from an oxidized Si wafer. The PDMS substrate is oxygen plasma cleaned prior to transfer to improve its adhesion to the nanotubes. Figure 4(b) shows a topographic AFM image of the same nanotube taken after releasing the substrate. The image shows that the length of the nanotube has decreased by ~20%, and ~200 nm spatial period undulations in height are present that were absent in the original image. Similar undulations were also observed in several other samples we studied, with a typical spatial period ~100-200 nm. Fig. 4(c) shows line traces from the images, with the upper trace from Fig. 4(a), and the lower trace from Fig. 4(b) along the dotted lines. The undulations in the compressed tube have a characteristic height of ~1-2 nm.

We account for this behavior by considering that past a critical compression, materials undergo a buckling instability. In the presence of an elastic medium, such a buckling instability produces an undulatory distortion. This phenomenon was observed in two-dimensions for thin gold films[18] and Si[4] on PDMS. In our case, we must consider the one-dimensional rod-shaped geometry of the nanotubes. From elementary elasticity theory, the expected critical buckling force for a rod in contact with an elastic medium is given by

$$T_{cr} = \frac{\pi^2 EI}{L^2}\left(n^2 + \frac{\alpha L^4}{n^2 \pi^4 EI}\right), \qquad (1)$$

where $E$ is the Young's modulus of the nanotube, $I$ is its area moment of inertia, $L$ is the length, $\alpha$ is a parameter on the order of the Young's modulus of the PDMS substrate, and $n$ is the number of undulations, which minimizes the expression for $T_{cr}$. Minimizing $T_{cr}$ with respect to $n$ and using $\alpha = E_{PDMS}$ gives

$$\frac{L}{n} = \frac{\pi(EI)^{1/4}}{E_{PDMS}^{1/4}} \cong \pi R\left(\frac{E}{E_{PDMS}}\right)^{1/4}. \qquad (2)$$

Using $E=10^{12}$ Pa and $E_{PDMS}=2\times 10^7$ Pa[18] gives an undulation period of ~25 $D$, where $D$ is the tube diameter. Any error in α due to geometric factors is likely to be unimportant since it enters as a fourth root. For a typical nanotube diameter $D=2$ nm, this yields an undulation period of 50 nm. This value agrees with the order of magnitude of the observed undulation period, confirming that continuum elasticity theory accounts for the observed nanoscale buckling phenomenon.

Because of carbon nanotubes dependence of the band gap on strain, the non-uniform periodic strain produced by these modulations could enable the creation of a superlattice structure in which the local band gap is modulated with the same period as the mechanical undulations.[4] Such one-dimensional superlattice quantum-well structures may have novel electronic or optical properties[19].

In sum, we have used elastomeric PDMS substrates to strain nanotubes up to 20% elongation while performing simultaneous transport measurements. The devices are capable of sustaining strains as large as 20% while providing local strain transduction. In addition to potential applications, such as artificial skins, where soft materials are imbued with piezoresistivity by the carbon nanotubes and can sense local shape distortions, our results may provide a strategy for the facile alteration of nanotube electronic or optical properties, and the creation of superlattice structures within nanotubes. This may also enable, for example, other novel applications such as mechanically tunable light emitting diodes.

**Acknowledgements**

We thank Chun Ning Lau for helpful discussions. We acknowledge the support of ONR and Arrowhead Research Corporation.

**Figure Captions**

FIG. 1. Device fabrication, geometry, and operation. (a) Cr/Au Electrodes are evaporated through a shadow mask. A thin wire is placed on top of the shadow mask to produce a few μm gap between the electrodes, which are ~30 μm in width. Nanotubes are placed on the electrodes from a suspension in dimethylformamide. (b) AFM image of completed device. A patch of a nanotube network bridges the electrodes. The electrical transport characteristics are measured through attached source and drain electrodes. (c) Individual single-walled nanotube before (left) and after (right) the substrate is elongated. The changes in the nanotube dimensions in the *x* and *y* directions are indicated.

FIG. 2. Transport characteristics of nanotubes under strain. (a) Device with ~10 nanotubes bridging the gap. The resistance changes from ~50 MΩ to 1 GΩ under an applied strain of 10%. (b) Device resistance with several patches of percolating nanotube networks bridging the electrodes. The resistance increases with strain. The device is strained and released from 0-10%

and 10-20% to demonstrate the reproducibility of the resistance changes. (Inset) Device resistance vs. strain. The resistance changes by ~1% per percent of strain.

FIG. 3. Histogram of band gaps under 0%, 10%, and 20% elongation for an ensemble of randomly oriented nanotubes with random chirality and diameter distribution 0.8-1.1 appropriate for HiPco nanotubes.

FIG. 4. Effects of compression on an individual single-walled carbon nanotube. (a) Topographic AFM image of nanotube prior to compression. (b) Topographic AFM image of nanotube after ~20% compression. Undulations appear in the image. (c) Line traces from images in (a) and (b) Top trace: line trace along dotted line in (a) Bottom trace: line trace along dotted line in (b). The height profile after compression shows distinct undulations that were not present before compression, that have a characteristic spatial period of 100-200 nm for the samples studied.

Fig. 1

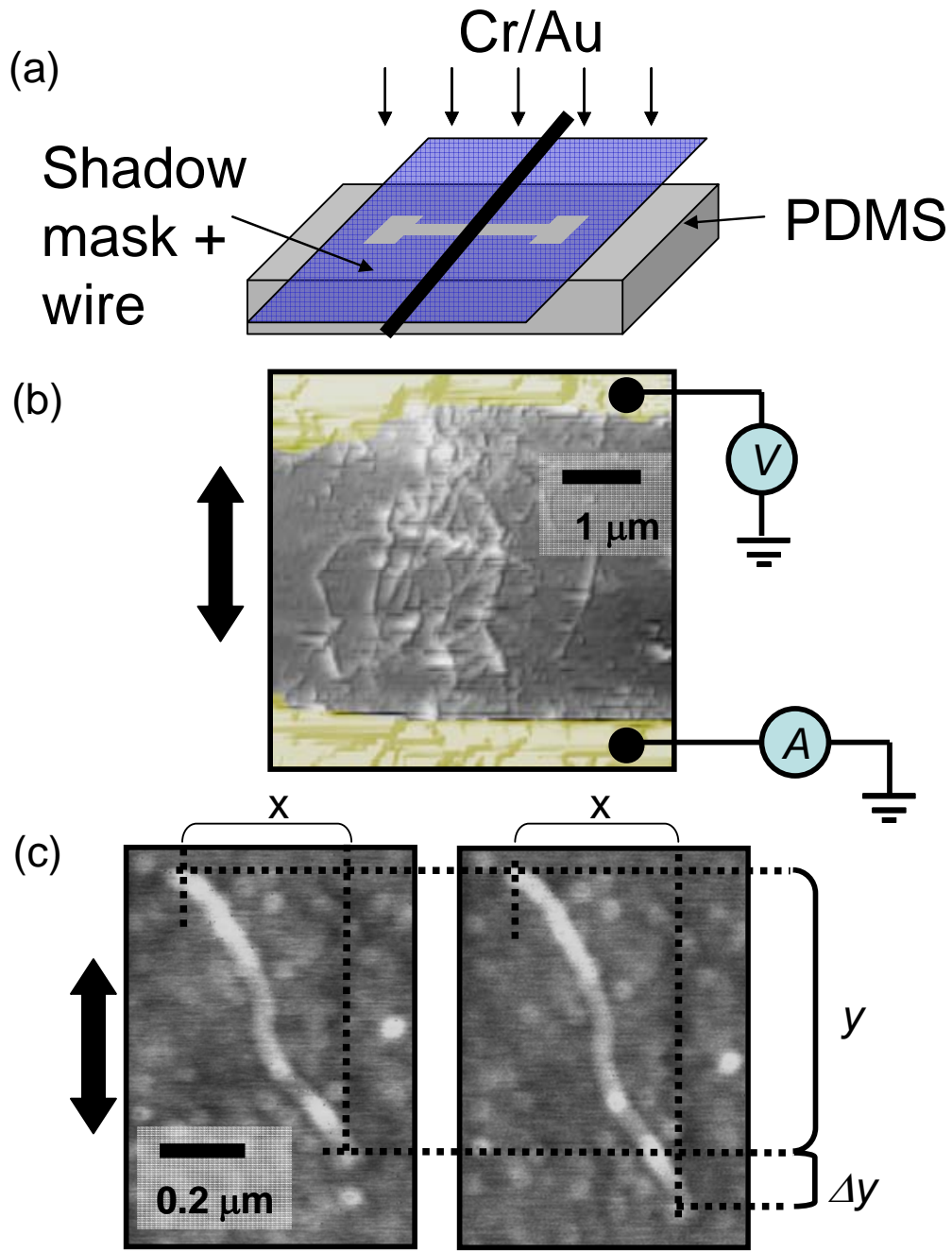

Fig. 2

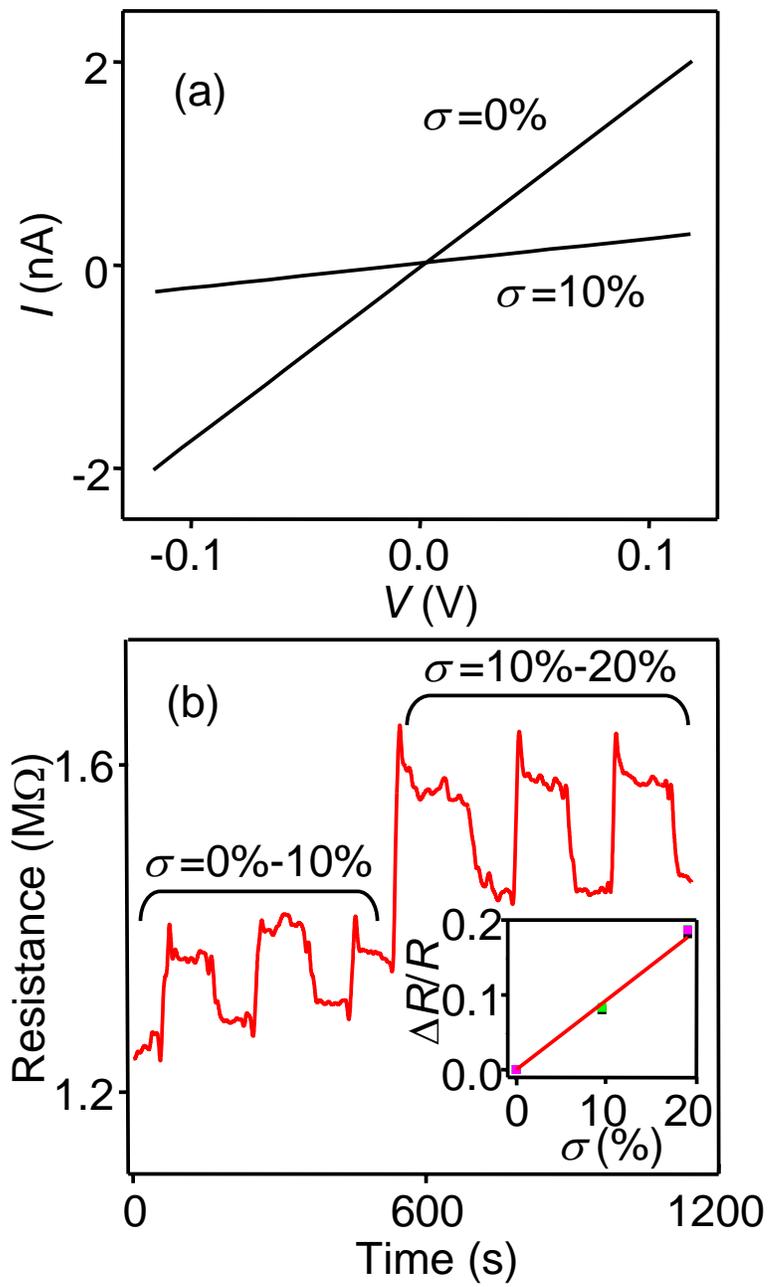

Fig. 3

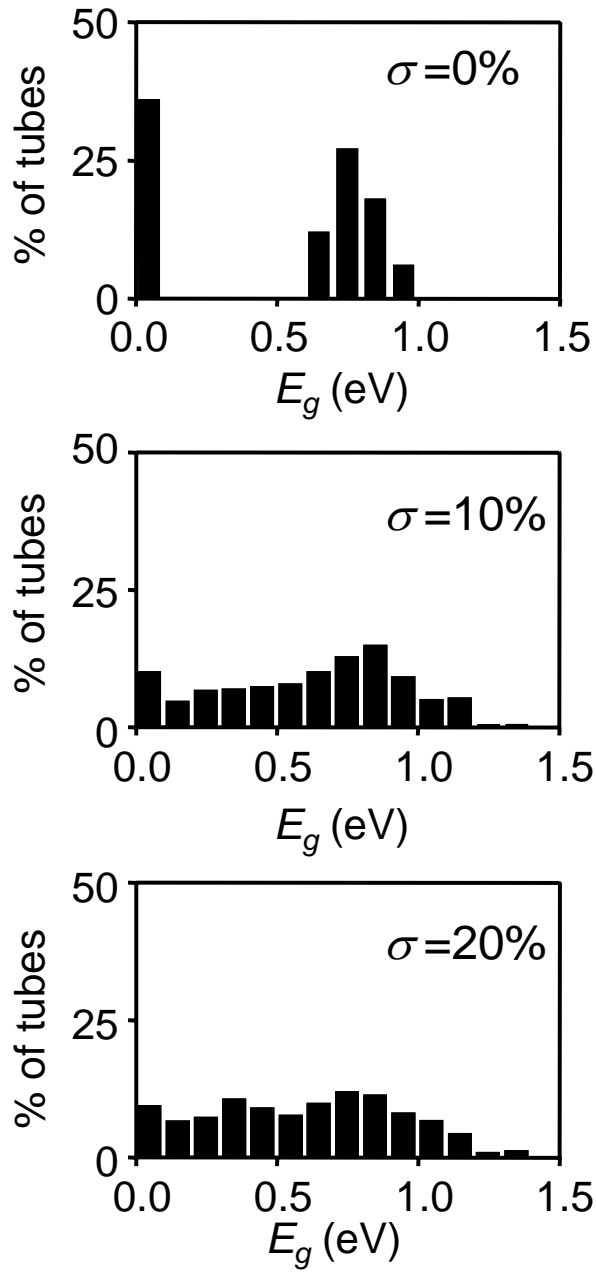

Fig. 4

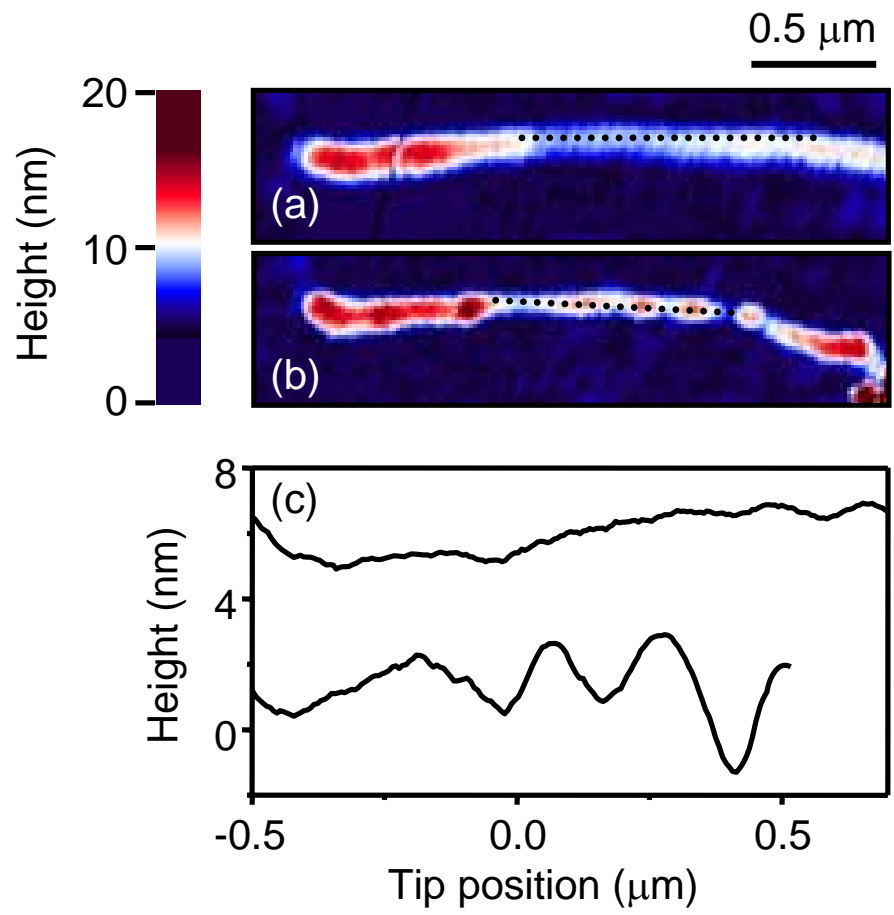